\documentclass[twoside,fleqn]{article}

\usepackage{npb}
\usepackage{epsfig}
\usepackage{axodraw}
\usepackage{amsfonts}

\newcommand{\eps}{\varepsilon}
\def\bq{\begin{eqnarray}}
\def\eq{\end{eqnarray}}
\def\l{\langle}
\def\r{\rangle}

\newcommand{\AmS}{{\protect\the\textfont2
  A\kern-.1667em\lower.5ex\hbox{M}\kern-.125emS}}

\title{Calculational techniques (not only) for single top production}

\author{Stefan Weinzierl\address{NIKHEF Theory Group,
        P.O. Box 41 882, NL - 1009 DB Amsterdam, The Netherlands}%
       }

\begin{document}

\begin{abstract}
A next-to-leading order calculation for single top production
including spin-dependent observables requires efficient techniques
for the calculation of the relevant loop amplitudes.
We discuss the adaption of dimensional regularization, 
the spinor helicity method and of tensor integral reduction algorithms
to these needs.
\end{abstract}

\maketitle

\section{Single top production}

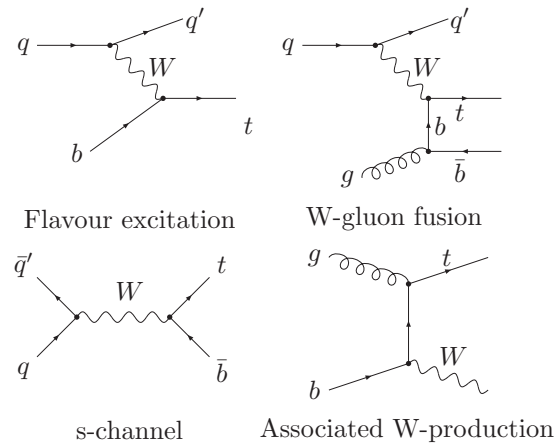
\begin{figure}
\begin{center}
\begin{picture}(400,130)(0,20)
\SetScale{0.5}
\ArrowLine(70,200)(125,240)
\Vertex(125,240){2}
\ArrowLine(125,240)(180,240)
\Photon(85,280)(125,240){4}{4}
\ArrowLine(30,280)(85,280)
\Vertex(85,280){2}
\ArrowLine(85,280)(140,300)
\Text(10,140)[]{$q$}
\Text(75,150)[]{$q'$}
\Text(62.5,132.5)[]{$W$}
\Text(95,110)[]{$t$}
\Text(30,100)[]{$b$}
\Text(50,75)[]{Flavour excitation}
\Gluon(275,180)(325,200){5}{4}
\Vertex(325,200){2}
\ArrowLine(380,200)(325,200)
\ArrowLine(325,200)(325,240)
\Vertex(325,240){2}
\ArrowLine(325,240)(380,240)
\Photon(285,280)(325,240){4}{4}
\ArrowLine(230,280)(285,280)
\Vertex(285,280){2}
\ArrowLine(285,280)(340,300)
\Text(110,140)[]{$q$}
\Text(175,150)[]{$q'$}
\Text(162.5,132.5)[]{$W$}
\Text(132.5,90)[]{$g$}
\Text(175,93)[]{$\bar{b}$}
\Text(175,115)[]{$t$}
\Text(167.5,110)[]{$b$}
\Text(150,75)[]{W-gluon fusion}
\ArrowLine(30,45)(60,75)
\ArrowLine(60,75)(30,105)
\Vertex(60,75){2}
\Photon(60,75)(130,75){4}{4}
\Vertex(130,75){2}
\ArrowLine(130,75)(160,105)
\ArrowLine(160,45)(130,75)
\Text(10,17.5)[]{$q$}
\Text(10,57.5)[]{$\bar{q}'$}
\Text(85,57.5)[]{$t$}
\Text(85,17.5)[]{$\bar{b}$}
\Text(50,47.5)[]{$W$}
\Text(50,-5)[]{s-channel}
\ArrowLine(250,20)(310,40)
\ArrowLine(310,40)(310,100)
\ArrowLine(310,100)(370,120)
\Vertex(310,40){2}
\Vertex(310,100){2}
\Gluon(250,120)(310,100){5}{4}
\Photon(310,40)(370,20){4}{4}
\Text(120,10)[]{$b$}
\Text(170,60)[]{$t$}
\Text(172.5,23)[]{$W$}
\Text(120,60)[]{$g$}
\Text(155,-5)[]{Associated W-production}
\end{picture} 
\end{center}
\caption{\label{fig1} Representative Feynman diagrams for single top 
production}
\end{figure}

The top quark, discovered 1995 at Fermilab, is 
special among the quarks due to its large mass:
Its lifetime is shorter than the characteristic
hadronization time scale and therefore 
top bound states do not have time to form.
From a calculational point of view this allows
an immediate application of perturbation theory 
to top physics. 
Up to now top quarks have only been produced
in pairs through the strong interaction.
With the upcoming Run II at the Tevatron and
later with LHC one expects to produce a top
quark also through an electroweak $Wtb$-vertex.
The principal production mechanisms for single
top production are flavour excitation,
W-gluon fusion, s-channel production and 
associated W-production.
Some representative Feynman diagrams are shown
in fig.~\ref{fig1}.
Single top production will be an essential input
for a direct measurement of the CKM matrix element
$V_{tb}$.
In addition, a discovery of non-standard charged-current top 
couplings might give a hint on new physics.
Since the electroweak decay of the top quark proceeds so rapidly before 
any hadronization effects can take place, the decay products of the 
top quark are correlated with the top quark spin.
Of particular importance will be the semileptonic decay $t\rightarrow  b \bar{l}\nu$,
since a detector signal corresponding to hadronic top decays will suffer
from large QCD backgrounds. 
Single top production, in which a top quark is produced
through a left-handed interaction, offers therefore 
an opportunity for polarisation studies.
Due to the short lifetime of the top quark,
production and decay of a top
quark should be considered together and the relevant processes
at the parton level are for example
$u + b \rightarrow b + \bar{l} + \nu + d$ for flavour excitation
or
$ u + g \rightarrow b + \bar{l} + \nu + d + \bar{b}$ for W-gluon fusion.
Here it is understood that we may replace the $(u,d)$-quark pair by
$(\bar{d},\bar{u})$, $(c,s)$ or $(\bar{s},\bar{c})$.
Of course most diagrams contributing to the two processes above do not
contain a top quark at all.
We performed a full leading-order calculation \cite{us} and compared the results
to the top narrow width approximation, to which only diagrams with an
intermediate top quark contribute. The top narrow width approximation
significantly reduces the number of diagrams.
Our results for the Tevatron are shown in table~\ref{num_res1}.
The first column gives the results from the full calculation, in the
second column we required in addition that the decay products of the top
reconstruct to within $20\;\mbox{GeV}$ to the top quark mass and
the third column gives the results for the narrow width approximation.
\begin{table*}
\begin{center}
\begin{tabular}{|c|c|c|c|} \hline
 Tevatron & $\sigma_{tot}$ & $\pm 20 \;\mbox{GeV}$ & narrow width \\ \hline
$W g$ & $15.0 \pm 0.4 \; \mbox{fb}$ & $14.3 \pm 0.3 \; \mbox{fb}$ & $14.5 \pm 0.1 \; \mbox{fb}$\\ \hline
$W b$ & $87 \pm 1 \; \mbox{fb}$ & $85 \pm 2 \; \mbox{fb}$ & $87 \pm 1 \; \mbox{fb}$\\ \hline
$q \bar{q}$ & $46 \pm 1 \; \mbox{fb}$ & $32.3 \pm 0.3 \; \mbox{fb}$ & $29.0 \pm 0.2 \; \mbox{fb}$\\ \hline
\end{tabular}
\end{center}
\caption{\label{num_res1} Numerical results for 
Tevatron at $2 \; \mbox{TeV}$ for W-gluon fusion ($Wg$), flavour excitation ($Wb$) and
s-channel production ($q\bar{q}$).}
\end{table*}
For W-gluon fusion we required three jets, two of them $b$-tagged, for
flavour-excitation two jets with one $b$-tag and for the s-channel
process we required two $b$-tagged jets.
Jets were defined by the hadronic $k_T$-algorithm \cite{kt}.
This exclusive signal definition allows us to distinguish the various
processes.
From table~\ref{num_res1} we see that the narrow width approximation
describes the cross section very well for W-gluon fusion and flavour
excitation. The approximation is less satisfactory for the s-channel
process. Here non-resonant terms seem to give a more sizeable
contribution.
We obtained similar results for the LHC.

\section{QCD corrections}

A leading-order calculation gives a rough description of the process
under consideration, but to reduce ambiguities due to the choice of
renormalization or factorisation scales a next-to-leading order (NLO)
calculation is required. 
Furthermore, if jets are defined an NLO-calculation models more accurately the internal
structure of a jet.
QCD corrections to the s-channel process and flavour excitation have been
considered in \cite{fe,sc}.
The calculation of QCD correction to W-gluon fusion is in progress \cite{pr}.
This calculation should give a more reliable prediction
on the $p_T$-spectrum of the $\bar{b}$-jet.
Since the top narrow width approximation works well for W-gluon fusion, 
it is sufficient to calculate QCD corrections in this
narrow width approximation. This simplifies the task considerably.
The calculation of the relevant loop amplitudes can in principle 
be done entirely with conventional
methods: One approach could be to calculate
\bq
2 \; \mbox{Re}\; A_{Born}^\ast A_{Loop}
\eq
and to use the Passarino-Veltman algorithm for the reduction of tensor integrals \cite{pas}.
One uses further the 't Hooft-Veltman prescription \cite{hooft} for $\gamma_5$, 
or, more efficiently, the reformulation according to S. Larin \cite{lar}.
Observables depending on spins can be treated within the spin density matrix formalism.
This solves the problem of calculating the one-loop amplitudes in principal, but leads
in practice to complicated expressions.
A more efficient approach is to calculate helicity amplitudes and to square them numerically.
Here the complexity grows linearly with the number of diagrams as opposed to a quadratic
increase within the conventional approach. The calculation of spin-dependent quantities is
trivial, since helicity amplitudes carry the complete spin information.
Efficient algorithm for the reduction of tensor loop integrals
rely on Fierz and Schouten identities and ``four-dimensional'' regularization
schemes like dimensional reduction \cite{DR} or the FDH-scheme \cite{FDH} 
are therefore favoured. (Of course these schemes are all variations
of dimensional regularization.)
Since weak interactions do not conserve parity,
$\gamma_5$ makes it appearance and one carefully has to avoid
inconsistencies inherent in some of the four-dimensional schemes.
We first review the spinor helicity method, before constructing
a regularization scheme adapted to $\gamma_5$, and comment finally on
reduction algorithms for tensor integrals.

\section{Spinor helicity method}

The spinor helicity method expresses the polarisation vectors for external
gluons of momentum $k$ in terms of two-component Weyl spinors $|p\pm \r$ and
$\l p \pm|$ as
\bq
\varepsilon^+_\mu = \frac{\l q- | \gamma_\mu | k- \r}{\sqrt{2}\l qk \r}, 
\varepsilon^-_\mu = \frac{\l q+ | \gamma_\mu | k+ \r}{\sqrt{2} [ kq ]}.
\eq
We have used the customary short-hand notation:
$\l ij \r = \l p_i - | p_j + \r$, $[ ij ] = \l p_i + | p_j - \r$.
Here $q$ is an arbitrary light-like
``reference momentum''.  The dependence on the choice of $q$ drops out
in gauge-invariant amplitudes.
In the narrow width approximation we treat the massive top quark as an
external state.
For a massive spinor we use \cite{mspin}
\bq
u(p,\pm) & = & \frac{1}{\sqrt{2 p q}} \left( p\!\!\!/ + m \right) | q \mp \r,
\nonumber \\
\bar{u}(p,\pm) & = & 
\frac{1}{\sqrt{2 p q}} \l q \mp | \left( p\!\!\!/ + m \right). 
\eq
Here, $p$
is the four-vector of the massive fermion with $p^2=m^2$ and
$p_0 > 0$, and $q$ is an arbitrary null vector with $q_0 > 0$.
It is easy to check that for these spinors the Dirac equations, 
orthogonality and completeness relations hold.
It will be advantageous to choose for $q$ the momentum of the charged
lepton from the top decay. This choice simplifies the factorisation
formula in the narrow width approximation.

\section{Variants of dimensional regularization}

We first give an overview of existing dimensional regularization schemes.
Naive dimensional regularization uses
\bq
\left\{ \gamma^\mu, \gamma^\nu \right\} = 2 g^{\mu\nu}, & &
\left\{ \gamma^\mu, \gamma_5 \right\} = 0.
\eq
In the 't Hooft-Veltman scheme \cite{hooft} a $D$-dimensional object is split between
a four-dimensional part and a $(-2\eps)$-dimensional part:
\bq
k_\mu^{(D)} = k_\mu^{(4)} + k_\mu^{(-2\eps)}, & & 
\gamma_\mu^{(D)} = \gamma_\mu^{(4)} + \gamma_\mu^{(-2\eps)}.
\eq
Here $\gamma_5$ is defined as a generic four-dimensional object:
\bq
\gamma_5 & = & \frac{i}{4!} \eps^{\mu\nu\rho\sigma} 
\gamma_\mu^{(4)} \gamma_\nu^{(4)} \gamma_\rho^{(4)} \gamma_\sigma^{(4)}.
\eq
As a consequence, $\gamma_5$ anticommutes with the first four Dirac matrices,
but commutes with the remaining ones:
\bq
\left\{ \gamma_\mu^{(4)}, \gamma_5 \right\} = 0, & &
\left[ \gamma_\mu^{(-2\eps)}, \gamma_5 \right] = 0.
\eq
Dimensional reduction \cite{DR} continues the momenta to $D<4$ dimensions,
but keeps spinors and vector fields in four dimensions.
\bq
\left\{ \gamma_\mu, \gamma_\nu \right\} & = & 2 g_{\mu\nu}^{(4)}
\eq
The $D$-dimensional metric tensor $g_{\mu\nu}^{(D)}$ acts also 
as a projection operator:
\bq
\label{proj}
g_\mu^{(D)\rho} g^{(4)}_{\rho\nu} & = & g_{\mu\nu}^{(D)}.
\eq
Naive dimensional regularization is not consistent
In that scheme one can derive an equation like
\bq
(D-4) \mbox{Tr}\; \gamma_\mu \gamma_\nu \gamma_\rho \gamma_\sigma \gamma_5 & = & 0.
\eq
At $D=4$ this equation permits the usual non-zero trace of $\gamma_5$ with
four other Dirac matrices. However, for $D\neq4$ we conclude that the trace
equals zero, and there is no smooth limit $D\rightarrow4$ which reproduces
the non-zero trace at $D=4$.
A similar equation can be derived in dimensional reduction, and
dimensional reduction is therefore also algebraically inconsistent.
Here the inconsistency can be related to the 
projection property eq.~\ref{proj}.
The 't Hooft-Veltman scheme is consistent. It should be mentioned that
the 't Hooft-Veltman scheme violates the Ward identity for the axial current.
This Ward identity relates the following three diagrams:
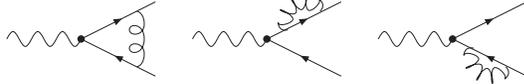
\begin{center}
\begin{picture}(300,30)(0,20)
\SetScale{0.7}
\ArrowLine(50,50)(90,70)
\ArrowLine(90,30)(50,50)
\Vertex(50,50){2}
\Photon(10,50)(50,50){4}{3}
\Gluon(80,65)(80,35){4}{2}
\ArrowLine(150,50)(190,70)
\ArrowLine(190,30)(150,50)
\Vertex(150,50){2}
\Photon(110,50)(150,50){4}{3}
\GlueArc(170,60)(10,210,30){4}{4}
\ArrowLine(250,50)(290,70)
\ArrowLine(290,30)(250,50)
\Vertex(250,50){2}
\Photon(210,50)(250,50){4}{3}
\GlueArc(270,40)(10,330,150){4}{4}
\end{picture}
\end{center}
The violation of the Ward identity is not a problem per se, but
one has to keep in mind, that one needs an additional finite
renormalization to restore the Ward identity.
To prove the (algebraic) consistency of a regularization scheme there are
two approaches:
The axiomatic approach of Breitenlohner and Maison \cite{BM}
takes $\gamma_\mu$, $\gamma_5$, $g_{\mu\nu}$ etc. as a set of abstract
symbols with a given set of relations and
shows that these relations are self-consistent.
The constructive approach of Wilson and Collins \cite{wil} tries
to find a represenation for each object in some 
(finite or infinite-dimensional)
vector space.
Here algebraic consistency is ensured by the fact that one has an explicit
representation.
In the following we will follow the constructive approach \cite{FD}.

\section{Constructing a regularization scheme}

We start with a simple example. Suppose, we are given the natural numbers
$1,2,3,...$ together with the rule how to add two numbers 
and we are asked to construct or to define negative numbers.
We can do that as follows: We consider pairs of natural numbers and call
two pairs, $(p,q)$ and $(k,l)$, equivalent, if
\bq
p+l & = & k+q.
\eq
An addition of these pairs is defined by
$(p,q)+(k,l)=(p+k,q+l)$. The set of equivalence classes yields then
the integer numbers. For example, the inverse of the class of $(p,q)$
is the class of $(q,p)$ and the neutral element is given by the class
of $(p,p)$.
Given the set of integer numbers and a rule of multiplication we can repeat
the exercise and construct the rational numbers.
A further example would be to consider pairs
$(V_i,V_j)$ and $(V_k,V_l)$ from a set of vector spaces 
$\{V_1,V_2,V_3,...\}$ and the equivalence relation
\bq
V_i \oplus V_l & \sim & V_k \oplus V_j.
\eq 
This construction is quite general and forms the basis of what mathematicians
call K - theory.
More formally, let $A$ be an abelian semi-group.
The $K$-functor associates to $A$ an abelian group $K(A)$, which is
constructed as follows:
Consider the equivalence relation on the set-theoretical product $A \times A$. We put
$(a,b) \sim (a',b')$ when there exists a $p \in A$ such that
\bq
a + b' + p & = & a' + b +p.
\eq
Then by definition $K(A) = A \times A / \sim$. 
We are now in the position to start the construction of our regularization
scheme.
Let ${\cal V} = \{V_1,V_2,...,V_i,...\}$ be a set of 
finite-dimensional vector spaces.
${\cal V}$ is an abelian semi-group with respect to the 
direct sum $\oplus$ and the tensor product $\otimes$.
We use Grothendieck's K-functor twice to construct the corresponding
abelian groups and
define the rank of an quadruple $\left[ (V_i,V_j), (V_k,V_l) \right]$
by
\bq
\mbox{rank} & = & \frac{\mbox{dim}\;V_i - \mbox{dim}\;V_j}{\mbox{dim}\;V_k - \mbox{dim}\;V_l}.
\eq
Note the difference between the ``rank'' and the sum of dimensions
$\mbox{dim}\;V_i+\mbox{dim}\;V_j+\mbox{dim}\;V_k+\mbox{dim}\;V_l$.
We can construct a set of quadruples such that their ranks form 
a dense subset of ${\mathbb R}$.
This construction can be extended to the complex case.
Furthermore we can work with vector space of even dimension only.
For a given 
quadruple of vector spaces of rank $r$ and whose sum of dimension is $2m$
we define the integration by
\bq
\lefteqn{\int d^rk f(k^2) = \frac{\pi^{r/2-m}}{\Gamma(r/2-m)} \int d^{2m}k } & & \nonumber \\
& &  
\cdot \int\limits_0^\infty dk_\perp^2 \left(k_\perp^2\right)^{r/2-m-1} f(k^2+k_\perp^2)
\eq
The integration is well-defined, e.g. does not depend on the 
chosen representative. A different choice of a quadruple with the
same rank $r$ would yield the same result.
In addition the definition satisfies the usual properties of an
integration, e.g.
linearity
\bq
\int d^rk (a f_1 + b f_2 ) = a \int d^rk f_1 + b \int d^rk f_2,
\eq
translation invariance
\bq
\int d^rk f(k+q) = \int d^rk f(k),
\eq
the scaling law
\bq
\int d^rk f(\lambda k) = \lambda^{-r} \int d^rk f(k)
\eq
and the normalization
\bq
\int d^rk \exp(-k^2) = \pi^{r/2}.
\eq
To define the Dirac algebra we consider
\bq
V_4 \oplus V_{-2\eps}, 
\eq
where
$V_4$ is a four-dimensional vector space and $V_{-2\eps}$ a quadruple of
rank $-2\eps$ and total dimension $2m$.
We recall that we can work with vector spaces of even dimension only.
We now have two options to define the Dirac algebra.
The first possibility is to define
Dirac matrices of dimensions
$2^{2+m} \times 2^{2+m}$ over $(V_4 \oplus V_{-2\eps})$.
With 
\bq
\gamma_5 & = & i \gamma^0 \gamma^1 \gamma^2 \gamma^3
\eq
this yields the 't Hooft - Veltman scheme.
The second option consists in 
defining Dirac matrices of dimensions $2^2 \times 2^2$ over
$V_4$ and of dimensions $2^m \times 2^m$ over $V_{-2\eps}$ separately.
This corresponds to a four-dimensional scheme.
By construction this scheme is free of algebraic inconsistencies.
Basically the distinction between the rank and the sum of dimensions 
avoids the pitfall of dimensional reduction.
The scheme is specified by a mapping of the Feynman rules from $V_4$
to $V_4 \oplus V_{-2\eps}$.
The Dirac algebra in the numerator
can effectively be performed in four dimensions.
For the loop momentum appearing in the numerator of loop integrals one has
\bq
k\!\!\!/ k\!\!\!/ & = & k_{(4-2\eps)}^2 - k_{(-2\eps)}^2
\eq
The first term on the r.h.s can cancel a propagator, whereas the second
term shifts effectively the rank of the loop integral 
from $4-2\eps$ to $6-2\eps$.
As in the 't Hooft-Veltman scheme, the four-dimensional
scheme violates Ward identities, which have to be restored
by finite renormalizations.

\section{Tensor integral reduction}

Loop integrals may involve the loop momentum in the numerator.
To reduce these tensor integrals to standard scalar integrals,
several algorithm exists like 
the Pasarino-Veltman algorithm  
based on Lorentz invariance \cite{PV},
the Feynman parameter space technique \cite{FP}
or the partial integration technique \cite{PI}.
These algorithm are general but have also some drawbacks:
The first two introduce Gram determinants in the denominator, whereas
the last one might give rise to spurious poles in $1/\eps$.
For most practical applications more efficient algorithm exists.
For the calculation of helicity amplitudes
loop momenta are usually sandwiched between spinors. Consider the 
example
\bq
\gamma^\mu \gamma^\nu \int \frac{d^Dk}{(2 \pi)^D} 
\frac{k_\mu k_\nu}{(k^2-m_1^2) ... ((k-q_n)^2-m_n^2)}
\eq
A tensor reduction according to Passarino-Veltman would result
in long intermediate expressions, but rewriting
\bq
\gamma^\mu \gamma^\nu k_\mu k_\nu & = & (k^2-m_1^2) + m_1^2
\eq
reduces the integral immediately. More systematically, one uses
the Fierz identity
\bq
\l A+ | \gamma_\mu | B+ \r \l C- | \gamma^\mu | D- \r = 2 \left[ AD \right]
\l CB \r,
\eq
the Schouten identity for spinors
\bq
\l AB \r \l CD \r & = & \l AD \r \l CB \r + \l AC \r \l BD \r,
\eq
as well as the Schouten identity for four-vectors
\bq
\lefteqn{
\eps(q_1,q_2,q_3,q_4) k_\mu = \sum\limits_{a=1}^4 \left(k \cdot q_a \right)
v^a_\mu, } & & \nonumber \\
& & v^a_\mu = \frac{1}{6} \eps^{abcd} \eps(\mu,q_b,q_c,q_d).
\eq
The algorithm proceeds in two steps: First one reduces tensor integrals
to integrals with at most one power of the loop momentum in the numerator 
\cite{pittau}:
If there at least two massless external legs, say
$p_1$ and $p_2$, one writes
\bq
k\!\!\!/ & = & \frac{1}{2p_1 \cdot p_2} \left[ 
\left( 2 k \cdot p_2 \right) p\!\!\!/_1 + \left( 2 k \cdot p_1 \right) p\!\!\!/_2 
\right. \nonumber \\
& & \left. - p\!\!\!/_1 k\!\!\!/ p\!\!\!/_2 
- p\!\!\!/_2 k\!\!\!/ p\!\!\!/_1
\right].
\eq
The first two terms will cancel propagators, whereas a product of the last two terms can be rewritten as
\bq
\lefteqn{
\langle 1- | k\!\!\!/ | 2- \rangle \langle 2- | k\!\!\!/ | 1- \rangle = \left( 2 k \cdot p_1 \right) \left( 2 k \cdot p_2 \right) } & & \nonumber \\
& & - \left( 2 p_1 \cdot p_2 \right) k_{(4-2\eps)}^2 + \left( 2 p_1 \cdot p_2 \right) k_{(-2\eps)}^2, \nonumber \\
\lefteqn{
\langle 1- | k\!\!\!/ | 2- \rangle \langle 2+ | k\!\!\!/ | 1+ \rangle = \frac{1}{\langle 2- | 3 | 1- \rangle} } & & \nonumber \\
& & \cdot \left[ \left( 2 p_1 \cdot p_3 \right) \left( 2 p_2 \cdot k \right) \langle 1- | k\!\!\!/ | 2- \rangle \right. \nonumber \\
& & \left. -\left( 2 p_1 \cdot k \right) \langle 1- | p\!\!\!/_2 k\!\!\!/ p\!\!\!/_3 | 2- \rangle \right. \nonumber \\
& & \left. +\left( 2 p_3 \cdot k \right) \langle 1- | p\!\!\!/_2 k\!\!\!/ p\!\!\!/_1 | 2- \rangle \right. \nonumber \\
& & \left. - \langle 1- | p\!\!\!/_2 p\!\!\!/_3 p\!\!\!/_1 | 2- \rangle
\left( k_{(4-2\eps)}^2 - k_{(-2\eps)}^2 \right)
\right],
\eq 
In a second step the remaining vector integrals are reduced \cite{trace}.
For pentagon or higher point integrals one forms first the traces
$\mbox{Tr}_\pm k\!\!\!/ p\!\!\!/_1 p\!\!\!/_2 p\!\!\!/_3 p\!\!\!/_4 p\!\!\!/_5$
(here $\pm$ denotes the insertion of a helicity projection operator)
and uses then
\bq
\label{pent}
\lefteqn{
\mbox{Tr}_{\pm} \left( k\!\!\!/ p\!\!\!/_1 p\!\!\!/_2 p\!\!\!/_3 p\!\!\!/_4 p\!\!\!/_5 \right) = - \mbox{Tr}_{\mp}\left(p\!\!\!/_1 p\!\!\!/_2 p\!\!\!/_3 p\!\!\!/_4 \right) k_0^2 } & & \nonumber \\
& & - \frac{1}{2} \left( B_0 \pm \varepsilon(k-p_1,p_2,p_3,p_4) \right) k_0^2 \nonumber \\
& & + \frac{1}{2} \left( B_1 \pm \varepsilon(k,p_1+p_2,p_3,p_4) \right) k_1^2 \nonumber \\
& & - \frac{1}{2} \left( B_2 \pm \varepsilon(k,p_1,p_2+p_3,p_4) \right) k_2^2 \nonumber \\
& & + \frac{1}{2} \left( B_3 \pm \varepsilon(k,p_1,p_2,p_3+p_4) \right) k_3^2 \nonumber \\
& & - \frac{1}{2} \left( B_4 \pm \varepsilon(k,p_1,p_2,p_3) \right) k_4^2 \nonumber \\
& & - k_{(-2\eps)}^2  \left( \mbox{Tr}_\pm p\!\!\!/_1 p\!\!\!/_2 p\!\!\!/_3 p\!\!\!/_4 - \mbox{Tr}_\mp p\!\!\!/_1 p\!\!\!/_2 p\!\!\!/_3 p\!\!\!/_4 \right) \nonumber \\
\eq
The $p_i$ are the external momenta, the $k_i$ the momenta flowing 
through the loop propagators.
For readability we dropped a subscript $(4-2\eps)$ for the $k_i^2$.
$B_i$ depends only on the pinched integral under consideration and is given 
for box integrals by
\bq
B_i & = & s t - m_1^2 m_3^2 - m_2^2 m_4^2,
\eq
with $s$ and $t$ being Mandelstam variables and the $m_i$ denote the external masses. 
The derivation of eq.~\ref{pent} is based on a combination of 
the spinor algebra approach with
the dual vector approach \cite{jos}.
For box integrals one can use
\bq
\lefteqn{
\mbox{Tr}_{\pm}\left(k\!\!\!/_0 p\!\!\!/_1 p\!\!\!/_2 p\!\!\!/_3 \right) = \frac{1}{2} \left( B \pm \varepsilon(k_0,p_1,p_2,p_3) \right) } & &\nonumber \\
& & - \frac{1}{2} C_0 k_0^2  
 + \frac{1}{2} C_1 k_1^2 
 - \frac{1}{2} C_2 k_2^2  
 + \frac{1}{2} C_3 k_3^2  
\eq
where $C_i$ again depends only on the pinched integral:
\bq
C_i & = & m_1^2 + m_2^2 - m_3^2.
\eq
The $m_i$ are the external masses of the triangle.
This algorithm is most efficient, if most of the external and internal masses
are zero. It does not introduce Gram determinants in the denominator.
However, there might be cases where this algorithm cannot be applied
(for example if all external masses are non-zero). In that case one has to 
resort to the general algorithms mentioned above. 
Luckily, these cases are in practical calculations rather rare.

\section{Summary}

Single top production is not only of interest for a direct measurement
of $V_{tb}$ or for a search for signals of new physics, but offers
also the opportunity of polarisation studies.
In this article we discussed some technical issues relevant to 
a corresponding NLO-calculation.

\end{document}